\documentclass[prl,twocolumn,superscriptaddress,longbibliography]{revtex4-2}
\usepackage{amsthm, booktabs, color, mathtools, amsmath, amssymb, setspace, bbm, tensor, mathtools, placeins, graphicx, wrapfig, float, verbatim, xcolor, physics}
\usepackage[normalem]{ulem}
\usepackage{stmaryrd} 
\usepackage{units}
\usepackage{algorithm2e}
\usepackage{amssymb}
\usepackage{dsfont}

\newcommand{\prlsection}[1]{{\em {#1}.---~}}

\newcommand{\CQT}{Centre for Quantum Technologies, National University of Singapore, 3 Science Drive 2, Singapore 117543.\looseness=-1}

\definecolor{THc}{rgb}{0.9,0.3,0.2}

\renewcommand{\eqref}[1]{Eq.~(\ref{#1})} 

\graphicspath{{./00_figures/}}

\def\app#1#2{%
  \mathrel{%
    \setbox0=\hbox{$#1\sim$}%
    \setbox2=\hbox{%
      \rlap{\hbox{$#1\propto$}}%
      \lower1.1\ht0\box0%
    }%
    \raise0.25\ht2\box2%
  }%
}

\ifx\proof\undefined
\newenvironment{proof}[1][\protect\proofname]{\par
	\normalfont\topsep6\p@\@plus6\p@\relax
	\trivlist
	\itemindent\parindent
	\item[\hskip\labelsep\scshape #1]\ignorespaces
}{%
	\endtrivlist\@endpefalse
}
\providecommand{\proofname}{Proof}
\fi
\newtheorem{lem}{\protect\lemmaname}
\newtheorem{claim}{\protect\claimname}

\providecommand{\factname}{Fact}
\providecommand{\theoremname}{Theorem}
\providecommand{\claimname}{Claim}
\providecommand{\lemmaname}{Lemma}

\setcitestyle{sort&compress,numbers}
\usepackage[unicode=true, breaklinks=false, pdfborder={0 0 1}, backref=false, colorlinks=true, linkcolor=blue, citecolor=blue]{hyperref}

\begin{document}

\title{Resonance-Suppression Principle for Prethermalization Beyond Periodic Driving}

\author{Jian Xian Sim}
\email{simjianxian@u.nus.edu} 
\affiliation{\CQT}

\begin{abstract}

Non-equilibrium dynamics of strongly and rapidly driven quantum many-body systems is poorly understood beyond periodic driving, where heating is exponentially slow in the drive frequency (Floquet Prethermalization). In contrast, non-periodic drives were found to exhibit widely different heating scalings with no unifying principle. This work identifies a resonance-suppression principle governing slow heating up to a prethermal lifetime $\tau_*$: When the drive's spectral arithmetic structure restricts multiphoton resonances, $\tau_*$ is controlled by low-frequency spectral suppression. The principle distinguishes (i) Single-photon suppression, quantified by a low-frequency suppression law $f(\Omega)$ for the drive's Fourier Transform weight near $\Omega=0$, from (ii) Multi-photon suppression, where nested commutators remain controlled if exceptional arithmetic structure satisfies a subadditive property. Remarkably, if multi-photon suppression holds, $\tau_*$ scaling with drive speed $\lambda$ is governed by $f(\Omega)$. This law of $\tau_*$ is found through a small-divisor mechanism in this work's iterative rotating frame scheme. Multi-photon suppression breakdown separates $\lambda$-scaling of $\tau_*$ in linear response and non-perturbative theory, shown by a case study of Quasi-Floquet driving. The principle is applied to (i) Resolve inconsistencies in literature on non-periodic driving, and (ii) Provide design principles for engineering prethermal phases of matter in programmable quantum simulators, exemplified by new non-periodic `Factorial' drives with tunable $\tau_*$.

\end{abstract}

\maketitle

\prlsection{Introduction} A closed quantum many-body system subjected to time-dependent external driving generally undergoes unbounded energy absorption. The system usually approaches a locally featureless infinite-temperature steady state, which is said to be thermalized~\cite{Mori_2018}. However, non-equilibrium dynamics prior to thermalization demands a technically daunting analysis, and general theories are scarce. A well-known technique to characterise non-equilibrium properties is linear response theory (LRT), where the Kubo Formula is foundational~\cite{Kubo1957}. However, by assuming a perturbatively weak driving term $V(t)$ applied to a system initially in thermal equilibrium, LRT best describes physical systems gently nudged out-of-equilibrium. This work identifies a general mechanism for long-lived non-equilibrium stability in quantum many-body systems, under non-perturbatively strong driving. 

Far-from-equilibrium systems under strong driving produce physical phenomena beyond LRT. For instance, some strongly driven quantum systems undergo very slow heating until a time $\tau_*$, called `Prethermalization'. A notable example is high-frequency periodically driven (Floquet) systems~\cite{ADHHMagnus, ADHHFer, Kuwahara_2016Floquet, Mori_2016Floquet}, where $\tau_*$ grows exponentially with drive frequency. Floquet Prethermalization yields experimentally observed novel far-from-equilibrium phenomena, including prethermal time crystals~\cite{Else_2017, Kyprianidis_2021, moon2024experimentalobservationtimerondeau}, non-equilibrium symmetry-protected and symmetry-enriched topological phases~\cite{Else_Ho_Quasifloquet}. Prethermal dynamics are often analysed by high-frequency expansions which construct an effective Hamiltonian $H_{eff}$, describing a prethermal state $\rho_{local} \sim e^{-\beta H_{eff}}$~\cite{ADHHMagnus, ADHHFer, Kuwahara_2016Floquet, Mori_2016Floquet, Mori_2021ThueMorse}. $H_{eff}$ is local, and $\beta$ relates to initial state energy. 

Prethermal phenomena are observed across many experimental platforms, such as disordered dipolar spin ensembles in diamond~\cite{He_2023quasifloquet}, superconducting quantum processors~\cite{liu2025prethermalizationrandommultipolardriving}, nuclear spins in fluorapatite~\cite{Peng_2021}, and ultracold atoms~\cite{Rubio_Abadal_2020}. Technological applications of Floquet Prethermalization are starting to be proposed, such as prethermal time crystals for sensing~\cite{moon2024discretetimecrystalsensing}. However, applications are inevitably limited by considering only Floquet dynamics. 

In recent years, case-by-case analyses have emerged for prethermalization under strong non-periodic driving. Representative examples are $n$-random multipolar ($n$-RMD), Thue-Morse and smooth or non-smooth Quasi-Floquet driving~\cite{ADHHFer,ADHHMagnus,Mori_2021ThueMorse,Zhao_2021,Kuwahara_2016Floquet,Mori_2016Floquet,gallone2025}. A Quasi-Floquet drive has multiple incommensurate driving frequencies $\vec{\omega} = (\omega_1,...,\omega_m)$, such that $V(\vec{\omega}t) \coloneq \sum_{\vec{n} \in \mathbb{Z}^m} V_{\vec{n}} e^{i(\vec{n}.\vec{\omega})t}$, where $\vec{n}, \vec{\omega}$ are $m$-dimensional vectors~\cite{Else_Ho_Quasifloquet}. Thue-Morse drives are defined by step drives where $H_0$ and $H_1$ are applied in binary patterns $01101001...$~\cite{Zhao_2021} generated by an iterative substitution rule $0 \rightarrow 01, 1 \rightarrow 10$. 

These drives were presented in an unrelated fashion with diverse results for $\tau_*$'s scaling with driving speed $\lambda$, ranging from polynomial $\tau_*\sim \lambda^\alpha$ to stretched-exponential $\tau_* \sim e^{\lambda^{0.5}}$. Rapid advances in arbitrary waveform generation (AWG)~\cite{Bowler_2013,Bluvstein_2022,Martin_2023} enabled experimental implementation of these drives~\cite{moon2024experimentalobservationtimerondeau,He_2023quasifloquet,liu2025prethermalizationrandommultipolardriving}, and potentially other drives not yet discovered. Despite this prime opportunity to experimentally explore a broader landscape of drives, no unified understanding of the theoretical foundations for non-periodic driving was established.

Contrary to existing literature's emphasis on time-domain drive definitions, this work initiates a unified theory of prethermalization by identifying shared frequency-domain properties of non-periodic drives. A simple physical picture emerges in the frequency domain. Energy absorption by single-photon processes is controlled by the drive's Fourier Transform weight near zero frequency, described by a suppression law $f(\Omega)$ (see~\textit{Linear Response Theory)}. However, strong driving enables multi-photon processes that mix frequencies and can restore resonances even when single-photon absorption is suppressed. 
Slow heating therefore persists only when this low-frequency suppression remains stable under multi-photon mixing. 

Multi-photon stability is encoded by a subadditivity condition on the spectrum's arithmetic structure (see~\textit{Renormalization Scheme}), yielding an organizing principle for prethermalization beyond periodic driving. When multi-photon stability holds, $f(\Omega)$ fixes the $\lambda$-scaling of $\tau_*$, through a non-perturbative iterative rotating frame construction renormalizing the strong bare drive into a weak dressed drive. This iteration preserves locality and $f(\Omega)$ suppression in the final dressed frame (see~\textit{Non-Perturbative Theory} and~\textit{Renormalization Scheme}). 

Three representative `Prethermal Universality Classes' are listed in Table~\ref{tab:universalityclasses} with suppression law $f(\Omega)$, encompassing existing drives~\cite{ADHHFer,ADHHMagnus,Mori_2021ThueMorse,Zhao_2021,Kuwahara_2016Floquet,Mori_2016Floquet,gallone2025, Else_Ho_Quasifloquet} and a new `Factorial' construction (see~\textit{Applications})~\footnote{In principle, one can consider other suppression laws in between the three Prethermal Classes provided. This can be done explicitly, for example by appropriately tuning the decay rate of $|V_{\vec{n}}|$ w.r.t. $|\vec{n}|$ in a Quasi-Floquet drive such that $\tau_* \sim (\ln \lambda)^{\ln \lambda}$. But this is a rather artificial example, so for brevity I explicitly discuss only these three examples. It is possible that future drives discovered in a `naturally implementable' setting have $\tau_*$ scalings different from the three provided here. The necessary calculations follow the procedure of this work.}. Existing and new drives' $f(\Omega)$'s are derived in the SM~\cite{SuppMat}. Table~\ref{tab:universalityclasses} yields LRT predictions and non-perturbative bounds on effective Hamiltonians.

\prlsection{Setup}\label{sec::setup} Consider a finite lattice system $\Lambda$, not necessarily Euclidean, of spins with a local time-independent Hamiltonian $D^{(0)} = \sum_{Z\subset\Lambda} D^{(0)}_Z$, subject to a generically non-periodic driving $V(\lambda t) = \sum_{Z\subset\Lambda} V_{Z}(\lambda t)$ comprised of quasi-local driving terms. The full Hamiltonian is $H(t) = D^{(0)} + gV(\lambda t)$, where $g$ is the drive amplitude, not small in general. To be general, locality refers to graph-locality rather than geometric locality. \footnote{To bound the behaviour of local observables, we can strengthen the conditions to geometric locality and apply Lieb-Robinson Bounds.}

The drive can be converted into the frequency domain through a Fourier Transform~\footnote{For ease of explanation, a continuous frequency notation $\Omega$ is used. Mathematically, there are three types of spectral measures: Pure-point, absolutely continuous, singular continuous. In the Main Text, I avoid discussing such mathematical subtleties to focus on conveying the key physical ideas.}
\begin{equation}
V(\lambda t) \coloneq  \int_{\mathbb{R}} d\Omega \\\ \tilde{V}(\Omega)e^{i\lambda \Omega t} = 
\frac{1}{\lambda} \int_{\mathbb{R}} d\Omega \\\ \tilde{V}(\frac{\Omega}{\lambda})e^{i \Omega t}
\end{equation}
In order to express frequency components as $\lambda$-independent, I adopt the convention of having the Fourier Transform be $\tilde{V}(\frac{\Omega}{\lambda})$. The effect of increasing drive speed $\lambda$ is to stretch out the Fourier Transform $\tilde{V}(\frac{\Omega}{\lambda})$ along the horizontal axis. For example, in Floquet driving with an integer spectrum (coefficients of Fourier Series), the spectral gap widens as $\lambda$ increases. The asymptotic scaling of $\| \tilde{V}(\frac{\Omega}{\lambda})\|$ near the origin we represent by $f(\frac{\Omega}{\lambda})$, such that $\|\tilde{V}(\frac{\Omega}{\lambda})\| \sim f(\frac{\Omega}{\lambda})$ as $\Omega \rightarrow 0$. Physically, single-photon resonances are suppressed.

\begin{table}[t]
\caption{Prethermal universality classes characterized by suppression law $f(\Omega)$
and corresponding lower bounds on heating-time scaling $\tau_*^{LRT}, \tau_*^{NP}$ in LRT and non-perturbatively.}
\centering
\begin{tabular}{cccc}
\toprule
 & Poly & Quasipoly & Stretch-Expt \\
\midrule
$f(\Omega)$ 
& $|\Omega|^{b}$ 
& $e^{-|\ln \Omega|^{b}}$ 
& $e^{-1/|\Omega|^{b}}$ \\

$\tau_*^{\mathrm{LRT}}$ 
& $\lambda^{2b+1}$ 
& $e^{(\ln \lambda)^{b}}$ 
& $e^{\lambda^{b/(b+1)}}$ \\

$\tau_*^{\mathrm{NP}}$ 
& $\lambda^{b-1}$ 
& $e^{(\ln \lambda)^{b}}$ 
& $e^{\lambda^{b/(b+1)}}$ \\
\bottomrule
\end{tabular}
\label{tab:universalityclasses}
\end{table}

\prlsection{Linear Response Theory}\label{sec::linres} Laplace steepest-descent integration provides a unified linear response treatment of single-photon processes~\cite{BenderOrszag1999}. Given a local energy scale $J$ for performing a given local rearrangement, absorbing a quanta of energy $\Omega$ requires $O(\frac{\Omega}{J})$ local rearrangements. If $\Omega \gg J$, many local rearrangements are required for absorption, highly unfavourable and reflected in the decay of the Kubo Spectral Function $\sigma(\Omega) \sim e^{-\frac{\Omega}{J}}$ \cite{Abanin_2015linres}. Summing the absorption rate across different frequencies $\Omega$, with a perturbative driving amplitude $g$, applying the Kubo Formula predicts the heating rate below \cite{Kubo1957} as
\begin{equation}
    \begin{split}
        \frac{dE}{dt} \sim \int_{\mathbb{R}}d\Omega  \ |f(\frac{\Omega}{\lambda})|^2  e^{-\frac{\Omega}{J}} 
        =  \frac{g^2}{\lambda}  \int_{\mathbb{R}}d\Omega \ e^{2\ln[| f(\frac{\Omega}{\lambda})|] -\frac{\Omega}{J}}.
    \end{split}
\end{equation}
Letting $\phi(\Omega, \lambda) \coloneq 2\ln[| f(\frac{\Omega}{\lambda})|] -\frac{\Omega}{J}$, the rate of absorption at a fixed $\Omega,\lambda$ is $e^{-\phi(\Omega,\lambda)}$, maximised at some $\Omega_0 \coloneq \Omega_0(\lambda)$ , where $\phi'(\Omega_0) = 0$. Expanding $\phi$ in a Taylor Series about $\Omega = \Omega_0$, the integral is well approximated by keeping only the Gaussian term, $\frac{dE}{dT} \sim \frac{g^2}{\sqrt{\phi''(\Omega_0)}}e^{-\phi(\Omega_0)}$ (see SM~\cite{SuppMat}), thus we find $\tau_* \sim (\frac{dE}{dt})^{-1}$ (see Table~\ref{tab:universalityclasses}).

Observe that if the Fourier Spectrum $\tilde{V}(\Omega)$ can be approximated about the origin by a Taylor Series expansion, then the expansion's lowest non-vanishing power $\tilde{V}(\Omega) \sim \Omega^n$ asymptotically dominates the contribution to heating rate. By Table~\ref{tab:universalityclasses}, $\tau_* \sim \lambda^n$ can never be superpolynomial w.r.t. $\lambda$. Thus, those $f(\Omega)$ which provide superpolynomial $\lambda$-scaling of $\tau_*$, are precisely the infinitely-differentiable but non-analytic functions at $\Omega = 0$.

\prlsection{Non-Perturbative Theory}\label{sec::nonperturb} I construct a decomposition of the unitary time-evolution operator $U(t) \coloneq U^{(0)}(t) = \mathcal{T}(e^{-i \int_0^t dt' H(t')}) $ as
\begin{equation} \label{eqn:decomposition}
    U(t) = P_*(t) \mathcal{T}(e^{-i \int_0^t \ dt'[D_*+V_*(t')]}) \coloneq  P_*(t)  U_*(t)
\end{equation}
where $P_*(t) \coloneq e^{A^{(q_*-1)}(t)}...e^{A^{(0)}(t)}$ is a unitary rotating-frame composed of $q_*$ distinct unitaries, defined inductively later, called a Fer Expansion~\cite{Blanes_Casas_Oteo_Ros_2011, ADHHFer}. 
\begin{claim}
Suppose that $\| \tilde{V}(\Omega)\| \lesssim f(\Omega)$ as an upper envelope, with a spectral subadditivity condition Eqn~\ref{eqn:subadditivity}. By properties of $P_*,\ D_*, \ V_*$ in Eqn~\ref{eqn:decomposition} defined through recursion Eqn~\ref{equation:spectral}, $\tau_*$ obeys Table~\ref{tab:universalityclasses} for $\lambda$ sufficiently large.
\end{claim}
 Results on Quasi-Floquet and Factorial driving (see~\textit{Applications}) are rigorously shown in SM~\cite{SuppMat}. The SM introduces a related Discrete Fer Expansion for discontinuous step driving such as the Thue-Morse case, without rigorous bounds. Analytical and numerical evidence supports the same resonance-suppression mechanism. 

 The dressing $P_*(t)$ is near-identity, such that $D_*$ is near the time-averaged bare Hamiltonian $\langle H \rangle \coloneq \lim_{T\rightarrow \infty}\frac{1}{T} \int_0^T dt' H(t')$, and $V_*(t)$ is weak driving. More precisely,
\begin{equation}
    \|D_* -\langle H \rangle \|_{\kappa_*} \leq CJ \frac{J}{\lambda}, \quad
    \|V_* \|_{\kappa_*} \leq C' r^{q_*} \| V\|_{\kappa_0}\end{equation}
In this paper, $\|.\|$ is the operator norm, whereas $\| .\|_{\kappa}$ is a norm which quantifies operator amplitude, locality and resonance-suppression (see~\textit{Renormalization Scheme}). Crucially, $r \le 1/2$ as elaborated later, such that $\|V_*\|_{\kappa_*} \ll \|V\|_{\kappa_0} $. $J \coloneq \max\{  \|\langle H \rangle \|_{\kappa_0}\|, \|V\|_{\kappa_0}\}$ is a precise notion of local energy scale.

The main result is slow heating bounds of the form
\begin{equation}
    \frac{1}{J|\Lambda|} \| U^{\dagger}(t) \langle H \rangle U(t) - \langle H \rangle \| \leq \tilde{K} \frac{\| V_* \|_{\kappa_*}}{\| V \|_{\kappa}} t + O(J/\lambda)
\end{equation}
for all $t>0$. $\tilde{K}, C, C'$ are constants independent of the Hamiltonian $H(t)$, lattice geometry $\Lambda$ and drive speed $\lambda$. 

The heating bound says that the normalized energy density $\frac{\langle H \rangle}{J |\Lambda|}$ grows slowly up to a perturbative factor in $\frac{J}{\lambda}$. The energy density only grows significantly around $\tau_* \sim \tilde{K}^{-1}  \frac{\| V \|_{\kappa_0}}{\| V_* \|_{\kappa_*}} \sim r^{-q_*}$, depending on the ratio of the bare and dressed drive amplitude (see $q_*,\ r$ in Table~\ref{tab:smalldivisorproblem}). The SM~\cite{SuppMat} proves for the Polynomial case $q_* = b-1, \ r \sim \frac{1}{\lambda}$, for Quasipolynomial $q_* \sim (\ln \lambda)^{b-1}, \ r \sim \frac{1}{\lambda}$, for Stretched-Exponential $q_* \sim \lambda^{\frac{b}{b+1}}, \ r \sim \frac{1}{2}$. This furnishes the non-perturbative $\tau_*^{NP} \sim r^{-q_*}$ in Table~\ref{tab:universalityclasses}.

SM~\cite{SuppMat} further discusses dynamics of geometrically-local observables, if $\Lambda \subset  \mathbb{Z}^d$, in $d$-Euclidean dimensions.

To construct the decomposition of $U(t)$, define a superscript $O^{(q)}(t)$ for operators $O$, to indicate the $q$-th order of the iteration. Departing from the Schrodinger Equation $i \partial_t U^{(0)}(t) = H^{(0)}(t) U^{(0)}(t)$ in the initial bare frame, define inductively a rotating-frame transformation $U^{(q)}(t) = P^{(q)}(t) U^{(q+1)}(t)$, where $P^{(q)}(t) \coloneq e^{A^{(q)}(t)}$. $A^{(q)}(t)$ is an anti-Hermitian generator defined through the recursion equation $V^{(q)}(t) = i\partial_t A^{(q)}(t) $. 

Equivalently, by performing a Fourier Transform our recursion equation is defined by
\begin{equation}
\label{equation:spectral}
    \tilde{A}^{(q)}(\Omega) = -\frac{\tilde{V}^{(q)}(\Omega)}{\Omega}
\end{equation}
 with a dressed effective Hamiltonian $H^{(q+1)}(t)$ in the rotated frame, satisfying $i \partial_t U^{(q+1)}(t) = H^{(q+1)}(t) U^{(q+1)}(t)$, such that
\begin{equation}\label{eqn:dressedHamiltonian}
    \begin{split}
        H^{(q+1)}(t) = {P^{(q)}}^\dagger(t) H^{(q)}(t)P^{(q)}(t) - i {P^{(q)}}^\dagger(t) \partial_t P^{(q)}(t)  \\ \eqcolon D^{(q)} + W^{(q+1)}(t),
    \end{split} 
\end{equation}
where $W^{(q+1)}(t)$ has small amplitude (see End Matter and SM). To close the iteration at $q$-th order, we define our $(q+1)$-th dressed time-independent Hamiltonian $D^{(q+1)} = D^{(q)} + \langle W^{(q+1)}\rangle$, and zero-average time-dependent Hamiltonian $V^{(q+1)}(t) = W^{(q+1)}(t) - \langle W^{(q+1)}\rangle$. The above recursion is repeated until a truncation order $q_*$, explained later (see Table~\ref{tab:smalldivisorproblem}).

Eqn~\ref{equation:spectral} has a `small divisor problem' near $\Omega = 0$, indicating that the dressed drive has a weakened resonance-suppression. The iterative procedure trades reduction in drive strength for increased susceptibility to low-frequency resonances in the dressed frame. Eventually, resonance-suppression in the dressed frame is too weak and the iteration is terminated at a finite order $q_*$. Stronger suppression of $\tilde{V}^{(q)}(\Omega)$ near $\Omega = 0$ allows greater mitigation of the denominator's divergence, enabling more iterations performed, producing tinier dressed drive amplitude, ensuring longer $\tau_*$. Besides resonance, weakening locality in dressed frames also matter.

\prlsection{Renormalization Scheme} The following construction quantifies how under multi-photon stability, $\tau_*$ is controlled by the small divisor's renormalization of the dressed drive. I define `locality-resonance norms' $\| O \|_\kappa$ for time-dependent operators, by decomposing into quasi-local terms and respective frequency components $O(t) =\sum_{Z\subset \Lambda}\int d\Omega \ \tilde{O}_Z(\Omega)e^{i\Omega \lambda t}$. Let $|Z|$ be the size of the support of $|O_Z|$. Consider a `resonance-penalty' function
\begin{equation}
    p(\frac{\Omega}{\lambda}) \coloneq -\ln f(\frac{\Omega}{\lambda}).
\end{equation} For example, the stretched-exponential class defines a penalty $p(\frac{\Omega}{\lambda}) \coloneq (\frac{\lambda}{|\Omega|})^b \rightarrow \infty$ as $\Omega \rightarrow 0$. The locality-resonance norm schematically is
\begin{equation}
\| O\|_{\kappa} \coloneq \sup_{x \in \Lambda}\int_\mathbb{R} d\Omega  \sum_{Z \ni x} e^{\kappa[|Z| + p(\frac{\Omega}{\lambda}))]} \|\tilde{O}_Z(\frac{\Omega}{\lambda})\|.
\end{equation}
 In SM~\cite{SuppMat}, $d\Omega$ is formally integration over a measure $d\mu(\Omega)$, which sums over frequency indices for specific drive examples. The $\kappa$-norm aims to `penalize' frequency components near the origin which contribute most to single-photon resonances. 

 Given a generic $V(t)$ with a dense or continuous spectrum, the rotating frame $e^{A^{(0)}(t)}$ gives rise to nested-commutators in Eqn~\ref{eqn:dressedHamiltonian} convolving frequency components, reflecting how multi-photon processes restore resonances even under single-photon suppression. Two-photon simultaneous absorption and emission of frequencies $\Omega_1,\Omega_2$ may combine to a resonant scale
$\Omega_1+\Omega_2\sim J$ with
$p(J)\gg p(\Omega_1)+p(\Omega_2)$
violating subadditivity, inducing an unbounded $\kappa$-norm in dressed frames, thereby making prethermalization nongeneric. To exhibit prethermalization, a drive must possess the exceptional property that any two frequencies cannot add to form `too small' frequencies, captured by a subadditive property below.

By identifying an index map from frequency components $\Omega \rightarrow \mu(\Omega)$, $\mu$-index subadditivity is defined as
\begin{equation} \label{eqn:subadditivity}
    p(\mu(\Omega_1+...+\Omega_n)) \le p(\mu(\Omega_1)) + ....+p(\mu(\Omega_n)).
\end{equation}
For example, $\Omega = \vec{n}.\vec{\omega} \rightarrow \vec{n}$ for Quasi-Floquet driving, a `dyadic depth $d$' $\Omega \sim \frac{\lambda}{2^d} \rightarrow d$ for $n$-RMD/Thue-Morse driving, similarly `$p$-adic depth' for substitution rules mapping $0,1$ to length-$p$ strings~\footnote{A $p$-adic example was discussed by Mori et al.~\cite{Mori_2021ThueMorse}, but not using $p$-adic phrasing.}. The new `Factorial' drive has a `factorial depth' index $\Omega = \frac{\lambda}{k!} \rightarrow k$ (see~\textit{Applications}). Using such indices, I retain control of the $\kappa$-norm under rotating frames through a convergent cluster expansion bound (see End Matter and SM~\cite{SuppMat}).

Remarkably, if subadditivity of penalty holds, regardless of index details, single-photon suppression $f(\Omega)$ fixes heating rates through the `small divisor problem' of Equation~\ref{equation:spectral}, quantified by a $\kappa$-norm lemma.

\begin{lem} (Small Divisor Lemma)
Let $\kappa - \kappa' > 0$. Compute a `small divisor function', $h(\kappa - \kappa')$ below. I omit $\sup_{x \in \Lambda}$ in front for brevity.
\begin{equation}
\begin{split}
    \|A^{(q)}\|_{\kappa'} 
    &\coloneq  
       \sum_{Z \ni x} e^{\kappa' |Z|}
       \int_\mathbb{R} d\Omega\, e^{\kappa' p(\frac{\Omega}{\lambda})}
       \|\tilde{A}^{(q)}_Z(\frac{\Omega}{\lambda})\| \\  &=  
       \sum_{Z \ni x}  e^{\kappa' |Z|}
       \int_\mathbb{R} d\Omega\,  \frac{e^{-(\kappa - \kappa') p(\frac{\Omega}{\lambda})}}{|\Omega|} e^{\kappa p(\frac{\Omega}{\lambda})}
\|\tilde{V}^{(q)}_Z(\frac{\Omega}{\lambda})\|
        \\
    &\leq \sup_{\Omega \in \mathbb{R}} 
       \frac{e^{-(\kappa - \kappa') p(\frac{\Omega}{\lambda})}}{|\Omega|}
       \|V^{(q)}\|_\kappa \\
    &\eqcolon \frac{1}{\lambda} h(\kappa - \kappa') \|V^{(q)}\|_\kappa 
\end{split}
\end{equation}
\end{lem}
This lemma quantifies the severity of the `small-divisor problem', as measured in my $\kappa$-norm \footnote{For the Floquet case analyzed in \cite{ADHHFer}, it was found that $\|A\|_{\kappa'} \leq \frac{T}{2} \|V\|_\kappa$, there is no divergence. In fact, for the Floquet case, the lemma holds for $\kappa = \kappa'$. This highlights the absence of the `small divisor problem' unless one considers nonperiodically driven systems, upon which it becomes central.}. In the `small divisor problem', $h(\kappa - \kappa') \rightarrow \infty$ as $\kappa - \kappa' \rightarrow 0$ (See Table~\ref{tab:smalldivisorproblem}). The severity of the `small divisor problem' influences our choice of a sequence $\{\kappa_q\}_{0\le q \le q_*}$, severe singularities require faster drops in decay constants $\kappa_q - \kappa_{q+1}$. The Polynomial case's smallest allowed decay constant gaps are  $\kappa - \kappa' = \frac{1}{b}$, since each iteration of setting $\tilde{A}(\Omega) = -\frac{\tilde{V}(\Omega)}{\Omega}$ reduces suppression in the rotated frame by a factor $\Omega$, also numerically observed in SM~\cite{SuppMat}. The small divisor problem is extreme for Polynomially weak suppression, such that $\kappa - \kappa'$ cannot be arbitrarily small.

\begin{table}[t]
\caption{Rotating-frame transformation `small divisor problem' quantified by allowed decay gaps $\kappa - \kappa'$, small divisor function $h(\kappa-\kappa')$, 
Optimal truncation $q_*(\lambda)$, drive reduction ratio 
$r=\frac{\|V^{(q+1)}\|_{\kappa_{q+1}}}{\|V^{(q)}\|_{\kappa_q}}$, 
and beta function $\beta(\kappa)=\frac{d\kappa}{d\ln r}$.}
\centering
\resizebox{\columnwidth}{!}{%
\begin{tabular}{c|c|c|c}
\hline
 & Poly & Quasipoly & Stretch-Expt \\
\hline
$f(\Omega)$ & $\Omega^b$ & $e^{-|\ln\Omega|^b}$ & $e^{-1/|\Omega|^b}$ \\
$\kappa-\kappa'$ & $\ge 1/b$ & Any & Any \\
$h(\kappa-\kappa')$ & $1$ & $e^{1/(\kappa-\kappa')^{1/(b-1)}}$ & $(\kappa-\kappa')^{-1/b}$ \\
$q_*$ & $b-1$ & $(\ln\lambda)^{b-1}$ & $\lambda^{b/(b+1)}$ \\
$r$ & $1/\lambda$ & $1/\lambda$ & $1/2$ \\
$-\beta(\kappa)$ & $1/b$ & $[W((\lambda\kappa)^{1/(b-1)})]^{1-b}$ & $(1/\lambda\kappa)^{b/(b+1)}$ \\
\hline
\end{tabular}%
}
\label{tab:smalldivisorproblem}
\end{table}

Quasipolynomial suppression presents an essential singularity $h(\kappa - \kappa') \sim e^{\frac{1}{(\kappa - \kappa')^{\frac{1}{b-1}}}}$ which rapidly diverges as $\kappa - \kappa' \rightarrow 0$. In contrast, the Stretched-Exponential case has stronger suppression, resulting in a less severe polynomial singularity $h(\kappa - \kappa') \sim \frac{1}{(\kappa - \kappa')^{\frac{1}{b}}}$. The Stretched-Exponential case's gaps $\kappa_q - \kappa_{q+1}$ can be made smaller without having a `small divisor problem' as severe as the Quasipolynomial case, thus more iterations can be performed (see Table~\ref{tab:smalldivisorproblem} Optimal Truncation $q_*$). The driving amplitude in the dressed frame reduces by $r \coloneq \frac{\| V^{(q+1)} \|_{\kappa_{q+1}}}{\| V^{(q)} \|_{\kappa_q}} $ per iteration in Table~\ref{tab:smalldivisorproblem}). 

End Matter further elaborates on how to choose $\{ \kappa_q \}_{0 \le q \le q_*}$, by analysing the beta function $\beta(\kappa)$ of Table~\ref{tab:smalldivisorproblem} arising from small divisor renormalization, to reveal the scaling of $\tau_*^{NP}$ with $\lambda$.

\prlsection{Applications}\label{sec::application} Table~\ref{tab:universalityclasses} is now applied to resolve paradoxes in the literature and explain new Factorial drives.

For Quasi-Floquet drives, Else et al. showed that exponential decay $|V_{\vec n}|\sim e^{-|\vec n|}$ yields a stretched-exponential lifetime $\tau_*^{LRT},\tau_*^{NP} \sim e^{\lambda^{0.5}}$ via the penalty $p(\vec n)=|\vec n|$~\cite{Else_Ho_Quasifloquet}. In the SM, I show how smoother drives with faster Fourier Decay $|V_{\vec{n}}|$ display stronger $f(\Omega)$ suppression. One example is $|V_{\vec n}|\sim e^{-|\vec n|}$ for a two-tone drive, results in a soft spectral gap $f(\Omega) \sim e^{-\frac{1}{|\Omega|}}$, contained in the Stretched-Expt Class of Table~\ref{tab:universalityclasses}, for $b=1$. 

However, the experimentally realized two-tone drive by He et al.~\cite{He_2023quasifloquet}
\begin{equation}
    V_{QF}(t)\sim \sin(\lambda t)+\sin(\varphi\lambda t),\qquad
\varphi=\tfrac{\sqrt5-1}{2},
\end{equation}
has a hard spectral gap in its bare Fourier support. LRT would therefore suggest the stronger scaling $\tau_*^{LRT} \sim e^{\lambda}$. Non-perturbatively, however, the SM explains why heating cannot exceed $\tau_*^{NP} \sim e^{\lambda^{0.5}}$.

The mechanism is a breakdown of multiphoton suppression:
for generalized penalties $p(\vec n)=|\vec n|^{\alpha}$, subadditivity $|\vec{n}_1 + \vec{n}_2|^\alpha \le 
|\vec{n}_1|^\alpha+|\vec{n}_2|^\alpha$ holds \emph{only} for $0<\alpha\le1$. On the other hand, the hard gap of $V_{QF}(t)$ is the $\alpha \rightarrow \infty$ limit, arbitrarily fast superexponential decay w.r.t. $\vec{n}$.
Thus, Quasi-Floquet drives with a hard bare spectral gap have uncontrolled weakening of suppression under rotating frames when measured in $\kappa$-norms superexponentially decaying w.r.t. $|\vec{n}|$. This establishes multiphoton subadditivity as the cause of separating $\tau_*^{LRT}$ and $\tau_*^{NP}$, ensuring experimental data by~\cite{He_2023quasifloquet} is fitted appropriately.

Less smooth Quasi-Floquet drives span the remaining universality classes of Table~\ref{tab:universalityclasses}.
Polynomial and Quasipolynomial penalties,
$p(\vec n) \sim \ln|\vec n|$ or $(\ln|\vec n|)^2$,
satisfy subadditivity and imply respectively
$\tau_* \sim \mathrm{Poly}(\lambda)$ and
$\tau_* \sim e^{O[(\ln\lambda)^2]}$
(see SM~\cite{SuppMat}).
Hence, diverse scaling laws across Quasi-Floquet models arise from a single small-divisor renormalization mechanism.

Independently, Mori et al.~\cite{Mori_2021ThueMorse} proved that $\tau_*^{NP} \sim e^{O[(\ln\lambda)^2]}$ for the Thue-Morse drive, apparently contradicting the LRT expectations $\tau_*^{LRT} \sim e^\lambda$ by Zhao et al.~\cite {Zhao_2021}. The resolution: the hard spectral gap suggested by Zhao et al. is not true, Thue-Morse driving has a soft Quasipolynomial suppression near the origin~\cite{Baake_2019}~\footnote{It should be clearly stated that the result by Baake and Grimm consider the diffraction measure of the infinite Thue-Morse sequence, rigorously speaking a singular continuous measure. I neglect this mathematical difference for our physical exposition, expecting that the same mechanism discussed in this work applies, since the dyadic frequency index ensures subadditivity (see SM~\cite{SuppMat})}, analytically and numerically confirmed in SM. Thus, there is actually no disagreement between $\tau^{LRT}_*$ and $\tau^{NP}_*$.

More broadly, non-smoothly Quasi-Floquet driven systems and $n$-RMD drives both have $f(\Omega) \sim \mathrm{Poly}(\Omega)$ near the origin and subadditive penalties, yielding the unified predictions that $\tau_* \sim \mathrm{Poly}(\lambda)$~\cite{gallone2025, Mori_2021ThueMorse}. The Fibonacci drive~\cite{Dumitrescu_2018fibonacci}, being discontinuous Quasi-Floquet, also fits in this class~\footnote{However, as the SM~\cite{SuppMat} shows, Fibonacci drives only display $f(\Omega) \sim \Omega$ suppression near the origin, thus this work's results do not apply rigorously. It is however verified experimentally~\cite{He_2023quasifloquet} that such Quasi-Floquet step drives display a power law $\tau_* \sim \mathrm{Poly}(\lambda)$, and personal numerics not included in this work support such a power law scaling. A rigorous account of Fibonacci Prethermalization is beyond the scope of this work, although I conjecture that the physical mechanism is again resonance-suppression.}. The separation of $\tau_*^{LRT} \sim \lambda^{2b+1}, \ \tau_*^{NP} \sim \lambda^{b-1}$ is likely fundamental here -- a total loss of $f(\Omega) \sim \Omega^b$ suppression in dressed frames occurs after $b$ iterations. Rigorously, I show $q_* = b-1$ iterations in SM.

Finally, the principle further enables construction of new prethermal systems beyond Quasi-Floquet and step drives. For instance, consider a `Factorial' analytic non-periodic waveform~\footnote{This is the simplest toy model I am currently aware of beyond Quasi-Floquet and p-adic spectra, it is part of a broader class. I will discuss further examples and nuances in future work.}
\begin{equation}
    V^{F}(t) \coloneq \sum_{k=1}^{\infty} V^F_k \sin(\frac{\lambda}{k!}t)
\end{equation}
for $|V^F_k| \rightarrow 0$ sufficiently fast with $k \rightarrow \infty$, a faster $|V^F_k|$ decay producing greater $f(\Omega)$ suppression. 

 $V^F(t)$ has a spectrum $\Omega_k = \frac{1}{k!}$, where addition of frequencies generate any rational number in $\mathbb{Q}$, a dense set in $\mathbb{R}$, raising concerns of multi-photon resonances. Fortunately, multi-photon suppression arises from its `factorial depth' frequency index obeying ultra-subadditivity: adding two frequencies $\frac{1}{k_1!}, \ \frac{1}{k_2!}$ of depth $k_1,\ k_2$, yields a new depth $\le\max\{k_1, k_2\}$, ensuring subadditivity for any non-decreasing penalty $p(k)$~\footnote{See SM for detailed discussion. Ultra-subadditivity also holds for the $p$-adic spectra of $n$-RMD and Thue-Morse drives. Ultra-subadditivity of $p$-adic depth ensures subadditivity for penalties in the $n$-RMD/ Thue-Morse case as well.}.

Consequently, decay of $|V^F_k| \sim (k!)^{-b},  e^{-(\ln k!)^b}, e^{-(k!)^b}$ produces the full hierarchy of prethermal scalings $\tau_*^{NP} \sim \lambda^{b-1}, e^{(\ln\lambda)^b}, e^{\lambda^{\frac{b}{b+1}}}$ respectively as per Table~\ref{tab:universalityclasses}, via penalties $p(k) \sim b\ln(k!), [\ln(k!)]^2, (k!)^b$ (see SM). This demonstrates that engineered spectral arithmetic directly controls non-periodic prethermal lifetimes.
 
\prlsection{Conclusion}\label{sec::conclusion} The principle reframes the search for nonperiodic prethermal phases as a problem of spectral arithmetic design. In this sense, spectral arithmetic plays a role analogous to organizing principles such as symmetry and topology in many-body physics: arithmetic isolates structural constraints that enable far-from-equilibrium stability classes across disparate microscopic realizations.~\cite{ArnoldKozlovNeishtadt1997}~\footnote{Those familiar with classical mechanics may find the flavour of this paper similar to Kolmogorov-Arnold-Moser theory~\cite{ArnoldKozlovNeishtadt1997}. Indeed, I was influenced by their viewpoints on ergodicity-breaking. One objective of this paper is to show that long-lived ergodicity-breaking can arise in examples beyond KAM theory's `$\{ \vec{n}.\vec{\omega} \}$' frequency spectrum. The SM~\cite{SuppMat} will discuss how to understand analogues of the Diophantine condition of KAM Theory, in other cases such as dyadic and factorial spectra.}

The principle points towards systematic classification of long-lived nonequilibrium phases for prethermalization, including extensions for other slow relaxation phenomena such as classical prethermalization~\cite{Mori_2018ClassicalPrethermal} and disorder-induced localization under non-periodic driving~\cite{Dumitrescu_2018fibonacci,Long_2022,Tiwari_2024,Tiwari_2025,Zhao_2022}. 

\begin{acknowledgments}
\prlsection{Acknowledgments}
 I thank Takashi Mori for helpful discussions and physically insightful questions on a first draft, that motivated a closer examination of multiphoton processes and refining the theory accordingly. I thank S. Pilatowsky-Cameo, H. Chang, and ChatGPT for discussions.  I thank W. W. Ho for lectures on Floquet Prethermalization in NUS PC5221 that stimulated this work, and discussions over the years. This work was supported by Center for Quantum Technologies, National Quantum Scholarships Scheme in Singapore.
\end{acknowledgments}

\bibliography{references}

\section*{END MATTER}

\prlsection{Constructing the renormalization flow} 

Here I discuss the technical procedure of choosing decay constants $\{ \kappa_q \}_{ 0\le q \le q_*}$, provide physical interpretation in our context.

The choice of $\{ \kappa_q \}_{ 0\le q \le q_*}$ is found for each universality class by calculating $r_q \coloneq \frac{\| V^{(q+1)} \|_{\kappa_{q+1}}}{\| V^{(q)} \|_{\kappa_q}}$, analyzing it reveals the scaling of $q_*$ with $\lambda$. The analytical method is not unique, here I discuss a generalization of a `differential equation' mathematical technique pioneered by Else et al.~\cite{Else_Ho_Quasifloquet}. See SM~\cite{SuppMat} for another approach which provides tighter bounds.

First we quantify the weakened drive in dressed frames. $W^{(q)}(t)$ in the Main Text takes the form
\begin{equation}
    \begin{split}
        W^{(q+1)}(t) \coloneq  
        [\gamma_q(D^{(q)}) - V^{(q)}] 
        +[\gamma_q(V^{(q)}) - V^{(q)}] \\
        -[\alpha_q(V^{(q)}) - V^{(q)}]
    \end{split}
\end{equation}
where we defined a shorthand 
\begin{equation}
\begin{split}
        ad_{A^{(q)}} O \coloneq [A^{(q)}(t), O],
    \gamma_q(O) \coloneq e^{ad_{A^{(q)}}}O,  \\
    \alpha_q(O) \coloneq \int^1_0 ds e^{ad_{sA^{(q)}}}O.
\end{split}
\end{equation} 
The three bracketed terms will be bounded individually.

To bound the dressed drive $\| V^{(q+1)} \|_{\kappa_{q+1}}$, where $V^{(q+1)}(t)  = W^{(q+1)}(t) - \langle W^{(q+1)} \rangle $, besides the Small Divisor Lemma, we need to use a generalized ADHH Iteration Lemma~\cite{ADHHFer}, which must be analyzed through frequency labels $\mu = \vec{n}, d, k$ (see SM~\cite{SuppMat}). 
\begin{lem} (ADHH Iteration Lemma)
For any $\kappa' < \kappa$ satisfying $3\|A\|_\kappa \le \kappa-\kappa'$, one has
\begin{equation}
\| e^{\mathrm{ad}_A} O - O \|_{\kappa'} 
\le \frac{18}{\kappa'(\kappa-\kappa')} \|A\|_\kappa \|O\|_\kappa.
\end{equation}
\end{lem}
The above expression is a general combinatorial cluster expansion bound generalizing ADHH's Floquet bound in~\cite{ADHHFer}, valid only under frequency-index subadditivity. In the following discussion, since the ADHH lemma always takes the specific form above, the microscopic origin of subadditivity surprisingly plays no detailed role later on in heating rates! I explain why.

Combining the ADHH Lemma and the Small Divisor Lemma, we have
\begin{equation}\label{eqn:roughamplitudebound}
\begin{split}
    \| V^{(q+1)} \|_{\kappa_{\kappa_{q+1}}} \le 2\| W^{(q+1)} \|_{\kappa_{\kappa_{q+1}}} \\ \lesssim 18 \| A^{(q)}\|_{\kappa_q} \frac{(\| D^{(q)} \|_{\kappa_q} + 2\| V^{(q)} \|_{\kappa_q})}{\kappa_{q+1} (\kappa_q - \kappa_{q+1})} 
    \\ \lesssim 18 h(\kappa_q - \kappa_{q+1})\frac{\| V^{(q)} \|_{\kappa_q}}{\lambda} \frac{(\| D^{(q)} \|_{\kappa_q} + 2\| V^{(q)} \|_{\kappa_q})}{\kappa_{q+1} (\kappa_q - \kappa_{q+1})}
    \\ \lesssim  h(\kappa_q - \kappa_{q+1})\frac{\| V^{(q)} \|_{\kappa_q}}{\lambda} \frac{ 72 J}{\kappa_{q+1} (\kappa_q - \kappa_{q+1})}.
\end{split}
\end{equation}
In the above, we assumed that $\| V^{(q)} \|_{\kappa_q} \le J \coloneq \max\{\| D^{(0)} \|_{\kappa_0},\| V^{(0)} \|_{\kappa_0}\}$ and $\| D^{(q)} \|_{\kappa_q} \le 2J$, where $J$ is the precise definition of local energy scale in the bare frame, this claim is inductively justified in SM~\cite{SuppMat}. Physically, we expect this because the dressed frames $\| D^{(q)} \|_{\kappa_q}$ are intended to have local energy scales comparable to the bare frame $\| D^{(0)} \|_{\kappa_0}$.

The final line of Eqn~\ref{eqn:roughamplitudebound} is crucial, because we have included the effect of the `small divisor'. We rearrange to find \begin{equation}
    r_q \coloneq \frac{\| V^{(q+1)} \|_{\kappa_{q+1}}}{\| V^{(q)} \|_{\kappa_q}} \lesssim  \frac{h(\kappa_q - \kappa_{q+1})}{\lambda} \frac{ 72 J}{\kappa_{q+1} (\kappa_q - \kappa_{q+1})}.
\end{equation}
The $h(\kappa_q - \kappa_{q+1})$ term in the numerator reflects the contribution from the `small divisor problem'. The $(\kappa_q - \kappa_{q+1})$ in the denominator is from the cluster expansion dressing, this is the only source of divergence in the Floquet case.

Our aim is to choose a sequence $\{ \kappa_q \}_q$ such that each iteration reduces the $\kappa_q$-norm by at least $r_q \le \frac{1}{2}$, but possibly more. Keep in mind that the expansion defined by $\tilde{A}^{(q)}(\Omega) = -\frac{\tilde{V}^{(q)}(\Omega)}{\Omega}$ is fixed. On the other hand, our `human' choice of a $\{\kappa_q\}_q$ sequence can possibly be improved for tighter bounds. This historically occurred in the Floquet case~\cite{ADHHFer, Machado_2020}, achieved by Else et al. making the $\{ \kappa_q \}_q$'s a function of $\lambda$~\cite{Else_Ho_Quasifloquet}, which is also done here.

For a slowly decaying sequence of decay constants $\kappa \coloneq \kappa(q)$, approximating $\kappa_{q}-\kappa_{q+1} \approx -\frac{d\kappa}{dq}$, ignoring constant factors, we obtain a differential equation. Its solution provides a `reference curve' $\kappa(q)$ that keeps track of the weakening locality-resonance of our drive as we perform more rotating frames.
\begin{equation}
     r_q \sim O(1)\sim \frac{h[(-\frac{d\kappa}{dq})]}{\kappa (-\frac{d\kappa}{dq})} \frac{J}{\lambda}.
\end{equation}
$\kappa(q)$ is generally a concave-down function, the weakening of locality-resonance exacerbates with increasing $q$. To provide intuition on why concave-down, consider the Floquet case, often analyzed by the Floquet-Magnus Expansion~\cite{Mori_2016Floquet,Kuwahara_2016Floquet}. The combinatorial growth of terms becomes worse rapidly at high orders due to nested commutators, leading to an exacerbated weakening of locality (recall the full Floquet Hamiltonian $H_F$ is generally very non-local). Here, we have a generalized description of this phenomenon beyond Floquet-Magnus theory, by further accounting for the physical effect of weakening resonance-suppression in non-periodic driving.

\prlsection{Stretched-Exponential case} Explicitly, we obtain 
\begin{equation}
      1 = \frac{1}{\kappa (-\frac{d\kappa}{dq})^{\frac{b+1}{b}}} \frac{J}{\lambda} \iff \beta(\kappa) \coloneq \frac{d\kappa}{d\ln r}  \propto \frac{d\kappa}{dq} = -(\frac{J}{\lambda \kappa})^{\frac{b}{b+1}} ,
\end{equation}
as claimed in the Main Text Table~\ref{tab:smalldivisorproblem}. Upon rearranging and solving the differential equation with initial condition $q=1$ \footnote{We set $\kappa_1 \coloneq \kappa_0/2$ for some technical reasons discussed in~\cite{SuppMat}.}, we have 
\begin{equation}
\kappa_q = [(\kappa_1)^\mu - \frac{1}{\mu} (\frac{J}{\lambda})^{\frac{b}{b+1}} (q-1)]^{\frac{1}{\mu}}
\end{equation}
where $\mu \coloneq \frac{2b+1}{b+1}$. These results are a generalization of the functional forms found in~\cite{Machado_2020, Else_Ho_Quasifloquet}, for the Floquet and Quasi-Floquet cases respectively. 

In the Floquet limit $b \rightarrow \infty$, we have $1/\mu \rightarrow \frac{1}{2}$, such that $\kappa_q = \sqrt{(\kappa_1)^2 -  \frac{J}{2\lambda}q}$. Our choice mirrors the optimal choice of decay constants found in the Floquet case by~\cite{Machado_2020} (Appendix A2), which managed to saturate linear response theory bounds. On the other hand, by setting $b = \frac{1}{\gamma}$, where $\gamma$ is a Diophantine exponent, we recover the Quasi-Floquet form~\cite{Else_Ho_Quasifloquet}(Appendix H3).

Crucially, one can read off the scaling of $q_*$ with $\lambda$ by observing that when we fix $\kappa_{q_*} \coloneq \kappa_1/2 = \kappa_0/4$, upon making $q_*$ the subject of the equation, $q_* \sim (\frac{\lambda}{J})^{\frac{b}{b+1}}$.

\prlsection{Quasipolynomial case} We obtain 
\begin{equation}\label{quasipolynomialbetafunction}
      1 = \frac{e^{\frac{1}{(-\frac{d\kappa}{dq})^{1/(b-1)}}}}{\kappa (-\frac{d\kappa}{dq})} \frac{J}{\lambda} \iff \beta(\kappa) \sim -[W((\lambda \kappa)^\frac{1}{b-1})]^{-(b-1)}
\end{equation}
Rigorously extracting the scaling for the Quasipolynomial case is technical (see SM~\cite{SuppMat}), here I provide a qualitative argument. Starting from the LHS of Eqn~\ref{quasipolynomialbetafunction}, take the logarithm and rewrite it as 
\begin{equation}
 \frac{1}{(-\frac{d\kappa}{dq})^{1/(b-1)}} = \ln(\frac{J}{\lambda}) + \ln(\frac{1}{\kappa}) + \ln[(-\frac{d\kappa}{dq})^{1/(b-1)}]
\end{equation}
I argue that $\ln(-\frac{d\kappa}{dq})^{1/(b-1)}$ is negligible relative to $\frac{1}{(-\frac{d\kappa}{dq})^{1/(b-1)}}$, because for $\frac{d\kappa}{dq} \rightarrow 0^-$, the first is logarithmically divergent but the second is polynomially divergent. Furthermore, $\ln(\frac{1}{\kappa})$ also is negligible for large $\lambda$. We are left with
\begin{equation}
\begin{split}
    \frac{1}{(-\frac{d\kappa}{dq})^{1/(b-1)}}
    \approx \ln\!\left(\frac{J}{\lambda}\right)
    \implies
    \frac{1}{-\frac{d\kappa}{dq}}
    \approx 
    \Bigl[\ln\!\left(\frac{J}{\lambda}\right)\Bigr]^{b-1}
    \\
    \implies
    q \approx 
(\ln \frac{J}{\lambda})^{b-1}
    (\kappa_0 - \kappa_q)
    \implies
    q_* \sim
    O\!\left(
  [\ln\lambda]^{b-1},
    \right)
\end{split}
\end{equation}
where in the final line we set $\kappa_{q_*} \coloneq \kappa_0/4$. For $b= 2, q_* \sim \ln \lambda$, which agrees with the optimal truncation order of Mori et al.~\cite{Mori_2021ThueMorse}. The reduction of driving amplitude is weaker since we set $r_q \sim 1/2$, this bound is tightened in the SM~\cite{SuppMat} by a different technique, its key idea summarized as 
\begin{equation}
    \text{Minimize} \quad  \prod_{q=1}^{q_* - 1} r_q \quad \text{subject to constraint} \quad \kappa_1 - \kappa_* = \kappa_0/4,
\end{equation}

 achieved by a uniform spacing $\kappa_q-\kappa_{q+1} \coloneq const$.

\prlsection{Polynomial case} This method is distinct from the `differential equation' technique. Let $\kappa_0 = 1$, $\kappa_q - \kappa_{q+1} = \frac{1}{b}$, $h(\kappa_q - \kappa_{q+1}) = 1$, $r_q \sim \frac{1}{\lambda}$, if we let $\kappa_{q_*} = 1/b$,  such that iterating the transformation $b-1$ times, our final dressed drive amplitude scales as 
\begin{equation}
    \frac{\| V^{(q_*)} \|_{\kappa_{q_*}}}{\| V^{(0)} \|_{\kappa_0}}  \sim \frac{1}{\lambda^{b-1}}.
\end{equation}
Heuristically, the idea is that each iteration $A^{(q)} \coloneq -\frac{V^{(q)}(\Omega)}{\Omega}$ divides out a factor of $\Omega$, reducing the resonance-suppression in the dressed frame by a factor of $\Omega$, so iterations send $\Omega^b \rightarrow \Omega^{b-1} \rightarrow \Omega^{b-2} ... \rightarrow \Omega$. The SM provides numerics showing the reduction of spectral suppression in dressed frames, for $n$-RMD driving.

\end{document}